\documentclass[preprint,12pt]{elsarticle}

\usepackage{graphicx}
\usepackage{amssymb}
\usepackage{wasysym}
\usepackage{siunitx}
\usepackage{float}

\usepackage{lineno}

\journal{arXiv}

\begin{document}

\begin{frontmatter}


\title{Study of impact ionization coefficients in silicon with Low Gain Avalanche Diodes}


\author[a]{Esteban Curr\'{a}s Rivera\corref{cor1}}
\ead{esteban.curras.rivera@cern.ch}
\author[a]{Michael Moll\corref{cor1}}
\ead{michael.moll@cern.ch}
\cortext[cor1]{Corresponding author}

\address[a]{CERN, Organisation europ\'{e}nne pour la recherche nucl\'{e}aire, CH-1211 Geneva 23, Switzerland}



\begin{abstract}
Impact ionization in silicon devices has been extensively studied and several models for a quantitative description of the impact ionization coefficients have been proposed. We evaluate those models against gain measurements on Low Gain Avalanche diodes (LGADs) and derive new parameterizations for the impact ionization coefficients optimized to describe a large set of experimental data. We present pulsed IR-laser based gain measurements on 5 different types of  $50\mu m$-thick LGADs from two different producers (CNM and HPK) performed in a temperature range from  $-15\,^oC$ to $40\,^oC$. Detailed TCAD device models are conceived based on SIMS doping profiles measurements and tuning of the device models to measured C-V characteristics. Electric field profiles are extracted from the TCAD simulations and used as input to an optimization procedure (least squares fit) of the impact ionization model parameters to the experimental data. It is demonstrated that the new parameterizations give a good agreement between all measured data and TCAD simulations which is not achieved with the existing models. Finally, we provide an error analysis and compare the obtained values for the electron and hole impact ionization coefficients against existing models. 
\end{abstract}

\begin{keyword}
impact ionization \sep avalanche breakdown \sep electron multiplication \sep hole multiplication \sep gain \sep Low Gain Avalanche Diode.


\end{keyword}

\end{frontmatter}


\section{Introduction}
\label{S:Introduction}
\begin{figure}[!t]
\centering
\includegraphics[width=0.50\textwidth]{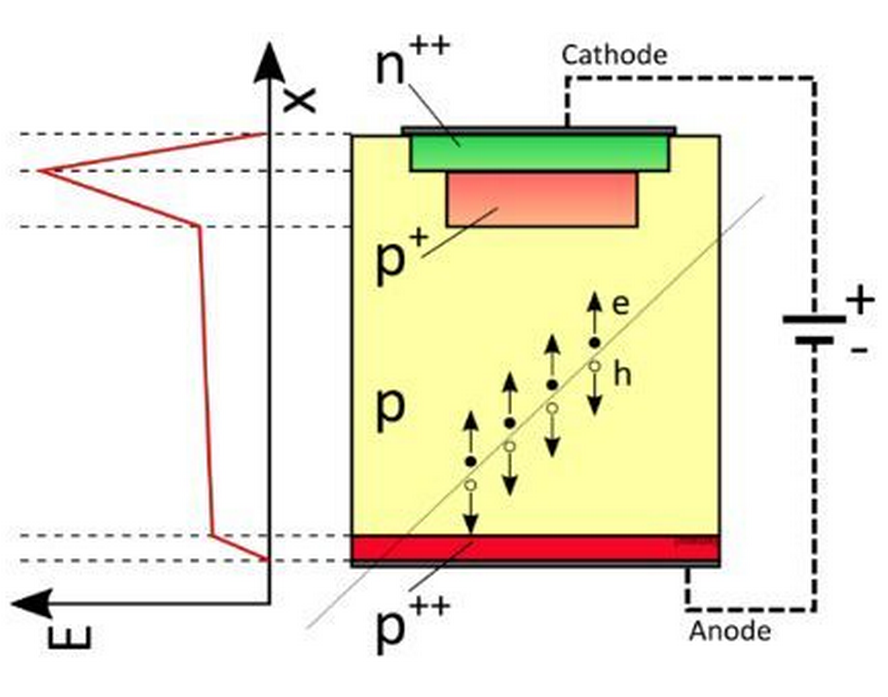}
\caption{A sketch of the cross-section of a pad-like LGAD with a charged particle passing through. The very highly doped implants of the front ($n^{++}$) and the back ($p^{++}$) electrode as well as the gain layer implant ($p^+$) within the high resistivity bulk ($p$) are shown. A qualitative profile of the electric field strength $E(x)$ is depicted too, where the peak is located in the gain layer region, in which the avalanche is taking place.} 
\label{Figure_1}
\centering
\end{figure}

Impact ionization in silicon arising from dedicated high field regions in devices, has been deeply studied and exploited for several sensor types over many decades. More recently, Low Gain Avalanche Diodes (LGADs) were developed within the RD50 collaboration to gain an internal signal amplification device for particle detection   \cite{PELLEGRINI201412}. They are implemented as $n^{++}-p^+-p$ avalanche diodes, where the highly-doped $p^{+}$ layer is added to create a very high electric field region. This electric field generates the avalanche multiplication of the primary electrons coming from the bulk of the device, creating additional electron-hole pairs in the gain layer region (GL). A schematic cross-section of a standard pad-like LGAD is shown in figure\,\ref{Figure_1}. The LGAD structure is designed to exhibit a moderate gain over a wide range of reverse bias voltage until the avalanche breakdown takes place. The LGAD technology is of high interest in the field of High Energy Physics (HEP) as a 4D tracking device \cite{Sadrozinski_2017}. It has been qualified for the use in the MIP Timing Detector of the CMS experiment and in the High-Granularity Timing Detector (HGTD) of the ATLAS experiment for the High Luminosity Large Hadron Collider (HL-LHC) operations \cite{CERN-LHCC-2017-027,CERN-LHCC-2018-023}.

To maximize the performance of LGADs, the device parameters have to be very carefully tuned to achieve the anticipated gain under the defined operational conditions
\cite{Pellegrini_2014, Wu_2020}. This requires reliable and precise impact ionization models to predict the gain in the design phase of the devices. There are several impact ionization models proposed in the bibliography \cite{MAES1990705, Massey, VANOVERSTRAETEN1970583,OKUTO1975161}. 
However, recent works on TCAD simulations of the gain of LGAD sensors have shown that the various impact ionization models deliver on the one hand significantly different results, and on the other hand exhibit often a limited agreement between simulated and experimental data  \cite{Mandurrino, Croci_2022, Yang_2021}. \\
This is partly due to the challenge to implement very precise device models that yield accurate electrical field configurations for the simulation. Especially the profiles of the implants that shape the high electric field region are not well known and can deviate between the anticipated (i.e. simulated profile) and the profile obtained during the fabrication process. The exponential dependence of the impact ionization coefficients on the electric field strength, and the exponential dependence of the multiplication factors on the length of the high electric field regions, result in big variations of the obtained gain with small changes in those device parameters.\\ 
Another source of uncertainty is coming from the measurements used to characterize the impact ionization. The measured gain is very sensitive to temperature and applied voltage calling for precise measurements of these parameters. More importantly, the measured gain can be reduced when high charge densities undergo amplification and create sufficiently high space charge densities to significantly reduce the electric field strength  \cite{CURRAS2022166530}. If not taken into account in both, measurement and modelling, this easily leads to errors in the impact ionization modelling. \\
All the above mentioned factors have to be carefully taken into account to properly model impact ionization in LGADs and are addressed in this paper. In the following, we describe the used samples and the employed measurement and simulation techniques (section \ref{S:samples}), followed by a detailed explanation of the methodology used to obtain a new set of parameters for the investigated impact ionization coefficients models (section \ref{S:Method}). Finally, we present the experimental data and the new optimized parameter sets with an in-depth error discussion (section \ref{S:Results}) and summarize our findings (section \ref{S:Summary}).

\section{Description and characterization of the samples}
\label{S:samples}
The samples used for this study were LGAD and PIN sensors produced by Hamamatsu Photonics (HPK) \cite{HPK} and Centro Nacional de Microelectr\'{o}nica (CNM-IMB) \cite{CNM}. The LGAD and PIN sensors from the respective producers differed only in the addition of the $p^+$-implant, i.e. the gain layer (GL), for LGADs.  The HPK samples were from the production run S10938-6130 (also called HPK prototype 2 or HPK2) produced on a wafer with a $50\,\mu m$ epitaxial layer on a $150\,\mu m$ thick low resistivity support wafer. The CNM samples were from the production run 12916, produced on a $50\,\mu m$ Float Zone wafer bonded to a $300\,\mu m$ low resistivity Czochralski wafer as support. The CNM LGADs were designed with a shallow gain layer doping profile and the same doping concentration for all devices investigated in the present work. The HPK ones have a deep gain layer doping profile and were produced in four different splits (S1, S2, S3, and S4) that differed in the doping concentration of the gain layer and consequently in the gain (see below).

\begin{figure}[tbh]
\centering
\includegraphics[width=0.95\textwidth]{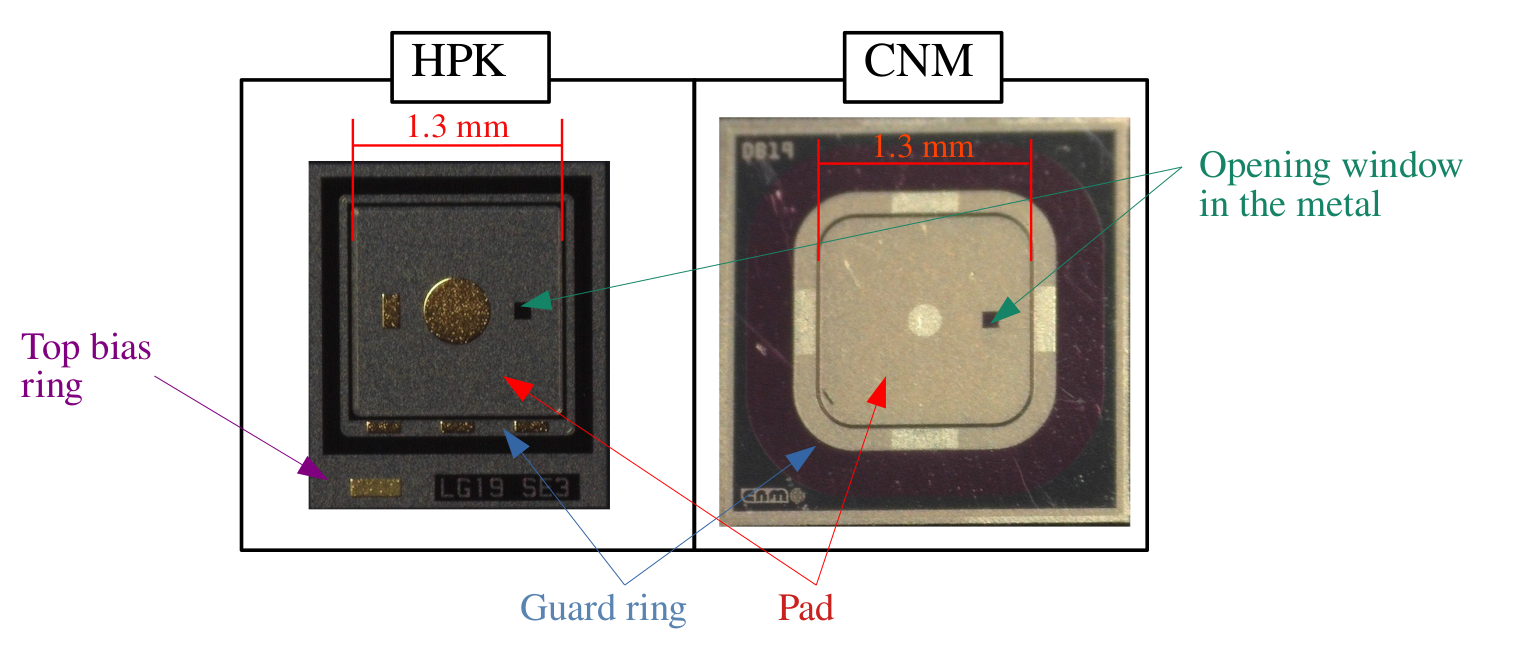}
\caption{Photographs (not scaled) of the two LGAD types studied in this work. The left picture shows a top view of an LGAD produced by HPK and the right one the top view of an LGAD produced by CNM. The corresponding PIN sensors of the two producers look identical.} 
\label{Figure:sample_photos}
\centering
\end{figure}

All samples have an active area of $1.3\times1.3\,mm^2$ and a guard ring structure surrounding the central pad. To allow laser illumination from the pad side (i.e. the front electrode), they have an opening window  of $100\times100\,\mu m^2$ in the metallization. In figure\,\ref{Figure:sample_photos}, two pictures of sensors studied in this work are shown, and in table\,\ref{table_1} the key parameters of the samples are listed. The depletion voltage of the PIN diodes $V_{dep}$, the depletion voltage of the gain layer of the LGAD $V_{gl}$, the breakdown voltage at $20\,^\circ C$ $V_{bd}(20\,^\circ C)$, the end capacitance of the LGAD sensors reached upon full depletion of the device $C_{end}$ and the active thickness of the sensors are given.

\begin{table}[!t]
\begin{center}
\begin{tabular}{| l | c | c | c | c | c |}
\hline
\textbf{Sample} & \textbf{V$_{dep}$} [V] & \textbf{V$_{gl}$} [V] & \textbf{V$_{bd}$\,(20$^o$C)} [V] & \textbf{C$_{end}$} [$p$F] & \textbf{d} [$\mu\,m$]\\
\hline
\hline
HPK-S1 & 7.2  & 54.5  & 145  & 3.6  & 48 \\
HPK-S2 & 7.2  & 53.5  & 168  & 3.6  & 48 \\
HPK-S3 & 7.2  & 51.2  & 212  & 3.6  & 48 \\
HPK-S4 & 7.2  & 50.7  & 235  & 3.6  & 48 \\
CNM    & 3.4  & 39.4  & 112  & 4.2  & 42 \\
\hline
\end{tabular}
\caption{Main parameters for the samples used in this work. The depletion voltage of the PIN ($V_{dep}$) and for the LGADs the gain layer depletion voltage ($V_{gl}$), the average breakdown voltage ($V_{bd}$) at 20$^o$C, the capacitance reached above full depletion ($C_{end}$) and the active thickness ($d$) are given.}
\label{table_1}
\end{center}
\end{table}

\subsection{Electrical characterization}
\label{SS:Electrical_characterization}

\begin{figure}[hbt!]
\centering
\includegraphics[width=0.75\columnwidth]{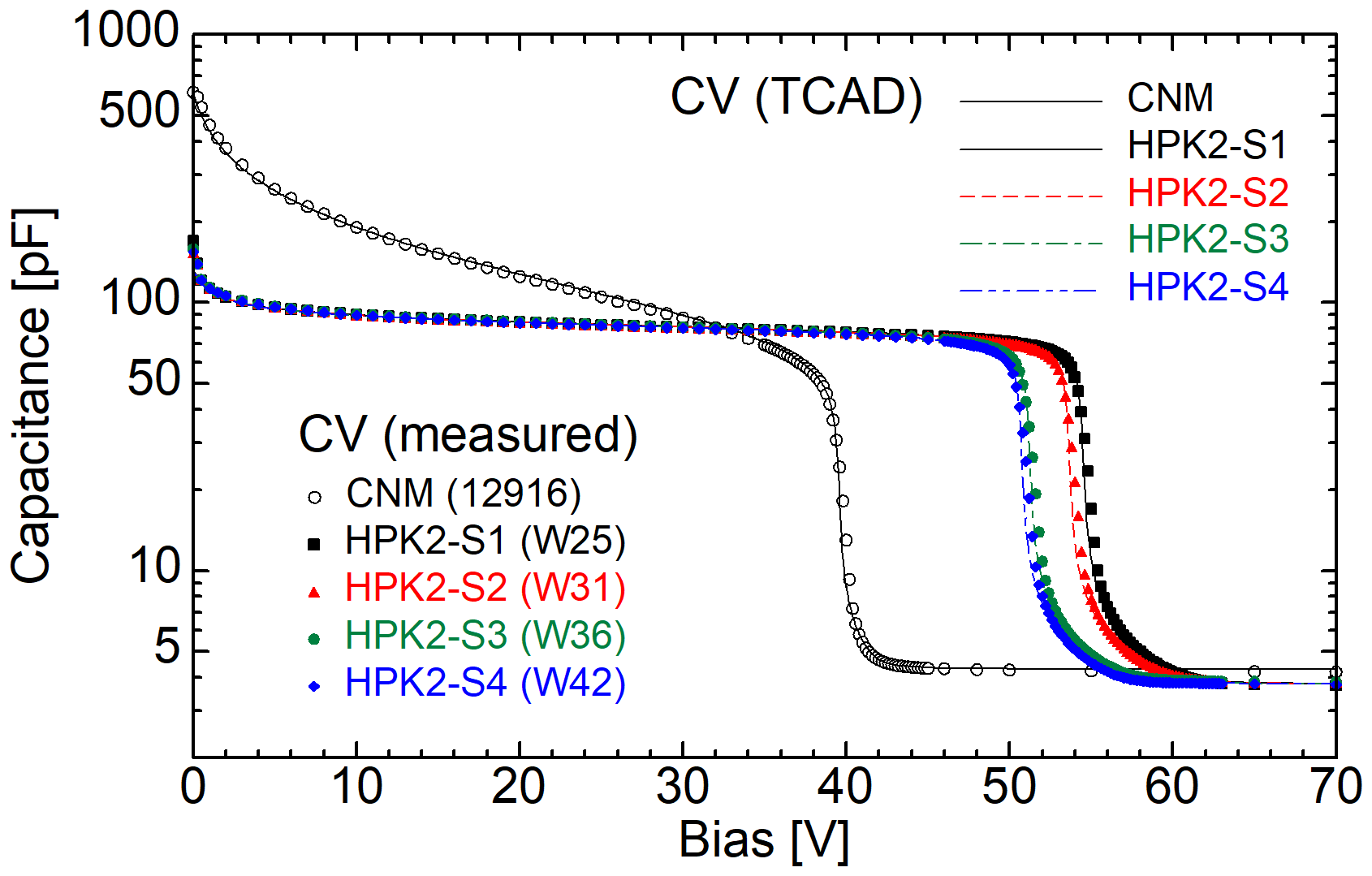}
\caption{Capacitance vs. voltage measurements (symbols) and corresponding TCAD simulations (lines) of the 5 LGAD types used for this work. The C-V curves were obtained at $20^\circ C$ with a measurement frequency of $1\,kHz$.} 
\label{Figure:CV}
\centering
\end{figure}

The electrical characterization was performed at the Solid State Detectors (SSD) lab at CERN. A probe station was used to measure the  leakage current and the capacitance as a function of the reverse bias voltage (I-V and C-V). The bare samples were placed directly on a temperature controlled chuck. During the electrical characterization the guard ring was connected to ground. All measurements were performed at $20\,^oC$ and the frequency in the LCR meter sinusoidal signal was set to $1\,kHz$ with an amplitude of $0.5\,V$. More details about the setup can be found in  \cite{CurrasRivera:2291517}. In  figure\,\ref{Figure:CV}, the C-V curves for the different LGAD types used in this work are shown. 

\begin{figure}[hbt!]
\centering
\includegraphics[width=0.75\columnwidth]{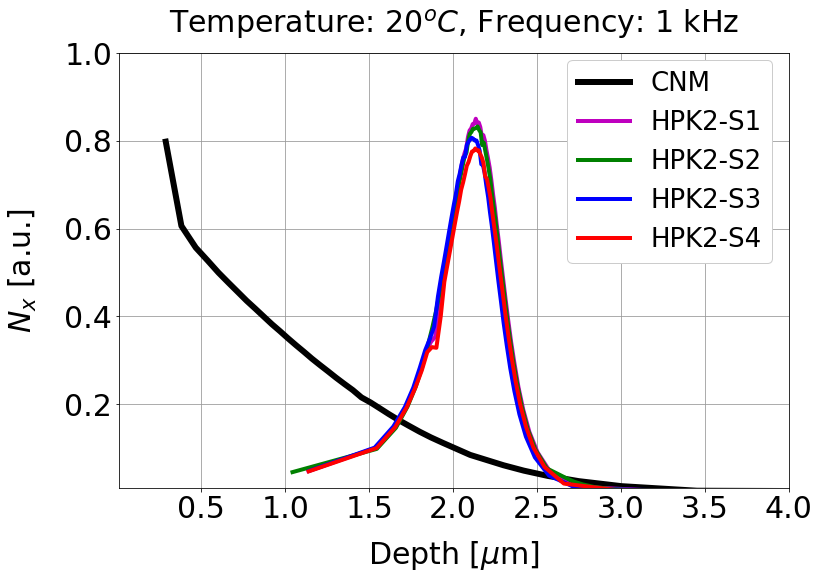}
\caption{Gain layer (GL) doping profiles extracted from the C-V measurements shown in figure \ref{Figure:CV} by employing equation \ref{Nx_eq}.} 
\label{Figure:Profile_CV}
\centering
\end{figure}

The GL doping profiles $N_x(x)$ as function of depth $x$ were extracted from the C-V curves using equation\,(\ref{Nx_eq})  \cite{schroder2015semiconductor}
\begin{equation}
\label{Nx_eq}
N_x(x) = \frac{C^3}{q \epsilon_0 \epsilon_r A^2 dC/dV}
\quad \text{with} \quad
x = \frac{\epsilon_0 \epsilon_r A}{C}
\end{equation}
where $C$ is the capacitance and $A$ the area of the sensor, $q$ the elementary charge, $\epsilon_0$ the permittivity of vacuum and $\epsilon_r$ the relative permittivity of silicon. The extracted doping profiles are given on the right hand side of figure\,\ref{Figure:Profile_CV}.
We can clearly distinguish the two different doping profiles used by the two producers: shallow by CNM and deep by HPK. It is however noted that the doping profile is extracted under the assumption of a constant area $A$ in equation\,(\ref{Nx_eq}). This assumption is a good estimate for the PIN diodes, but is only approximate for the small LGAD sensors with a wide Junction Termination Extension (JTE) as in the present case. The presence of the JTE and the lateral extension of the electric field, with growing bias voltage, allow only an approximate estimation of the doping profile employing equation\,(\ref{Nx_eq}). This is the reason why Secondary-Ion Mass Spectrometry (SIMS) measurements (see section \ref{SS:SIMS}), in combination with Technology Computer-Aided Design (TCAD) modelling (see section \ref{SS:TCAD_simulation}), were employed to extract the doping profiles and finally the electric field profiles.

\subsection{SIMS measurements}
\label{SS:SIMS}

\begin{figure}[hbt!]
\centering
\includegraphics[width=0.48\columnwidth]{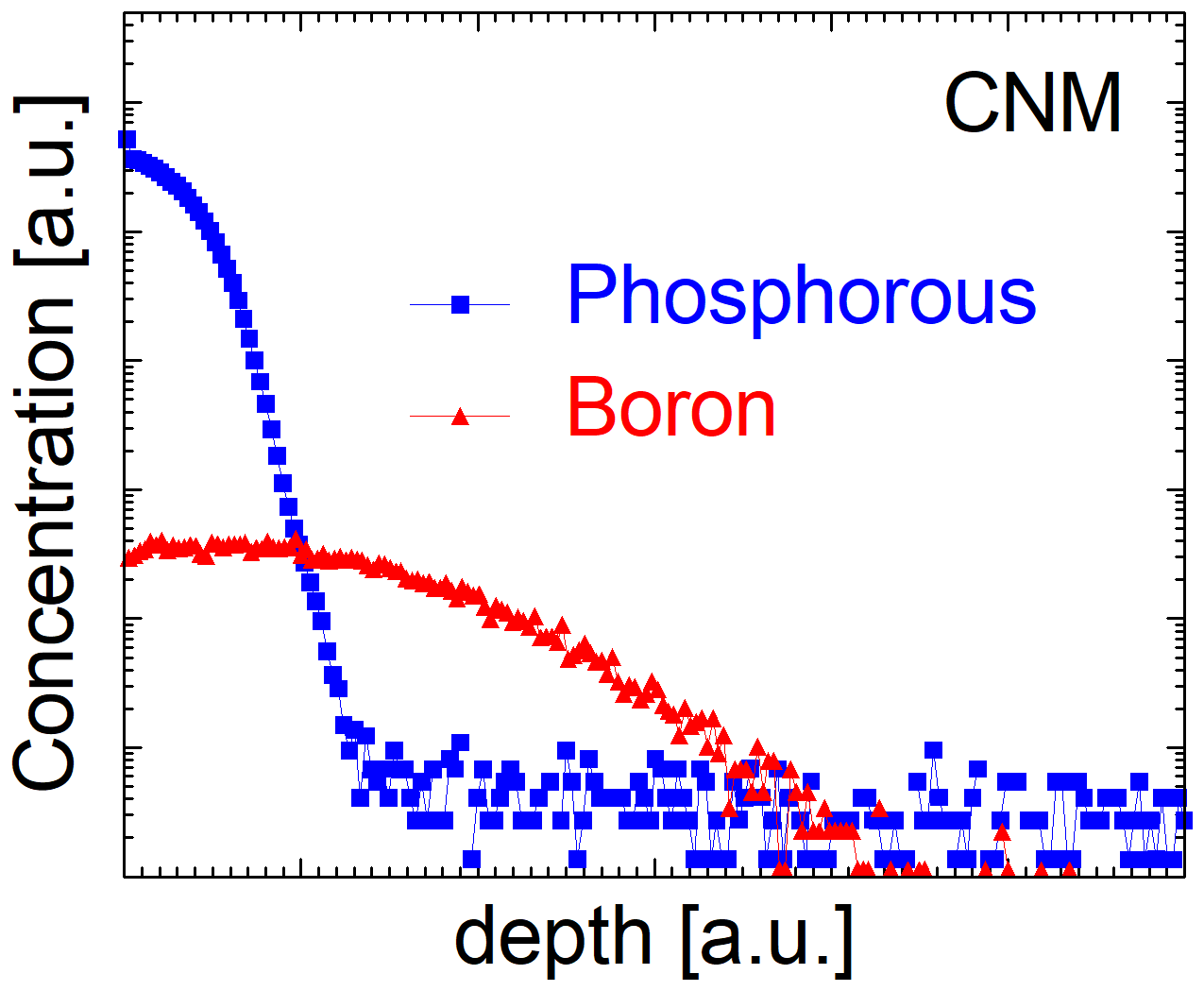}\hfil
\includegraphics[width=0.48\columnwidth]{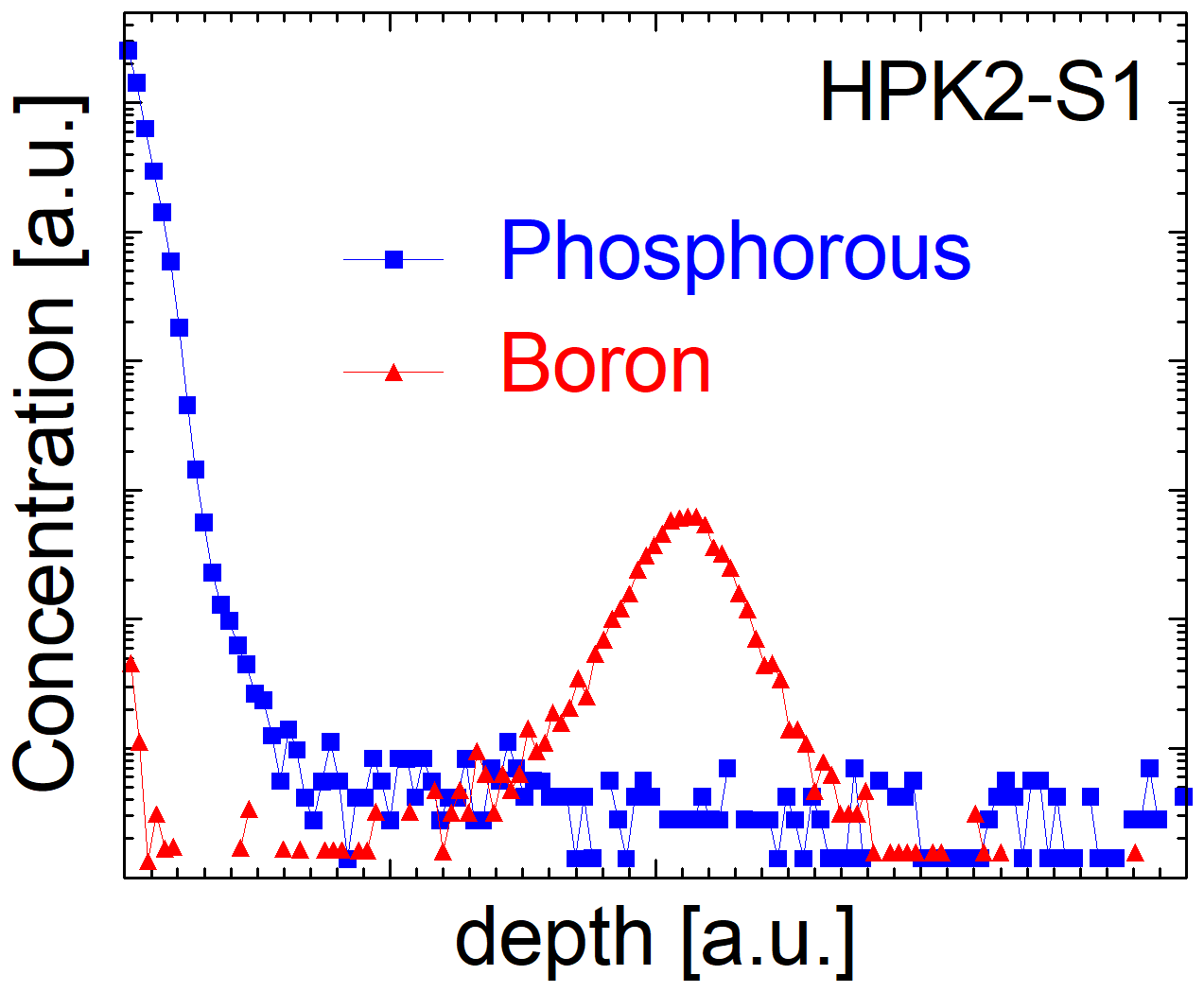}\hfil
\caption{Phosphorous and boron concentration profiles as measured with SIMS in the GL region of the CNM LGAD type (left) and the HPK2 S1 LGAD type (right).} 
\label{Figure_5}
\centering
\end{figure}

Secondary-Ion Mass Spectrometry (SIMS) analyses were performed on two types of samples at the SGS Institute Fresenius GmbH in Dresden, Germany \cite{SGS}. One CNM LGAD and one HPK2 S1 type LGAD were selected. The analysis of the data allowed to extract the phosphorus and boron concentrations in the gain layer region up to several microns in depth. In figure\,\ref{Figure_5}, the results of the measurements for the two sample types are shown in arbitrary units, with a linear $x$-axis and a logarithmic $y$-axis. In both cases, the boron concentration measured with SIMS, reflects the shape of the active boron profiles extracted from the C-V measurements given in figure \ref{Figure:CV}. Absolute values are not shown as requested by non-disclosure agreements with the sensor suppliers.

\subsection{Gain measurements with IR-laser}
\label{SS:Gain_IR}

The gain of the LGADs was measured using a pulsed IR-laser with a wavelength of $1060\,nm$ in the Transient Current Technique (TCT) setup of the SSD group at CERN. The IR-laser intensity was tuned to generate an equivalent charge of $\sim1\,MIP$ in the detectors and the spot size was $\diameter\approx80\,\mu m$. These two values were chosen to have a good SNR with low charge density inside the detector bulk to avoid any gain reduction up to a gain of 100. The gain was evaluated as the ratio between the charge measured in the LGAD and the charge measured in the PIN (after full depletion), both measured under the same conditions. The MIP equivalent charge calibration of the setup was determined by using a $^{90}S$r beta source on the same sensors. More details about the used TCT setup and the method employed to measure the gain with an IR-laser can be found in \cite{CURRAS2022166530}.\\

\begin{figure}[t]
\centering
\includegraphics[width=0.48\columnwidth]{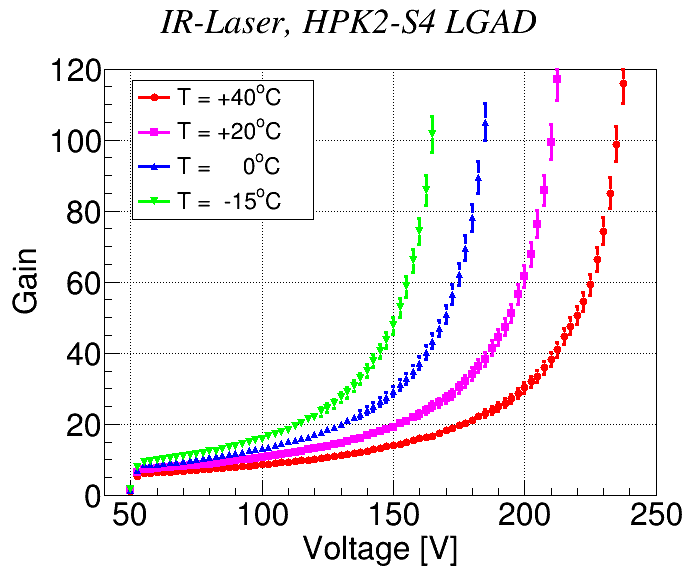}\hfil
\includegraphics[width=0.48\columnwidth]{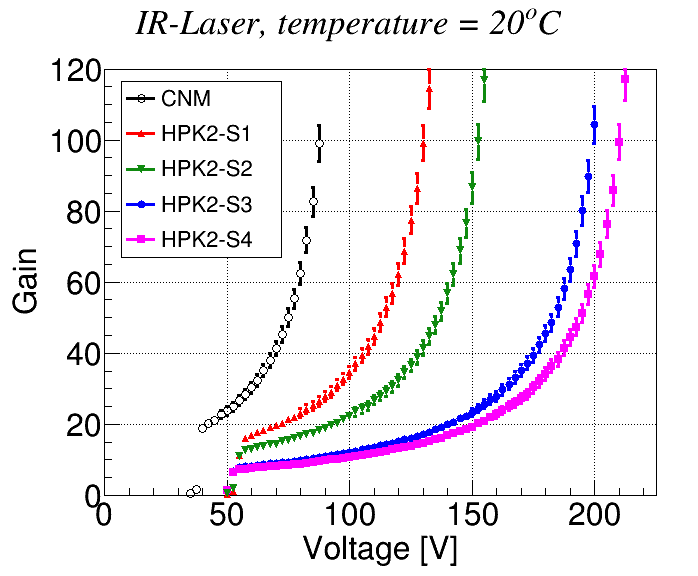}\hfil
\caption{On the left side: gain measured with the IR-Laser at different temperatures for the HPK2-S4 LGAD. On the right side: gain measure at $20\,^oC$ for all the LGAD types. The experimental error in the measurements is $\sim5\%$.} 
\label{Figure_4}
\centering
\end{figure}

The gain was measured at four different temperatures: $-15\,^oC$, $0\,^oC$, $20\,^oC$, and $40\,^oC$. 
Several samples were measured to estimate the error in the gain measurements, which was found to be in the order of $5\%$. 
This includes systematic errors in the measurements, fluctuation in the laser intensity, fluctuations in the temperature during measurements, and variation in the gain between "identical" LGADs (with the same expected gain by design). In figure\,\ref{Figure_4}, the gain at different temperatures for the HPK2 S4 LGAD is shown on the left side, and the gain measured at $20\,^oC$ for all used LGAD types on the right side.

\section{Methodology used for extracting impact ionization parameters}
\label{S:Method}

TCAD device simulations were performed in order to determine the electric field profiles as function of bias voltage in all LGAD sensors using the Sentaurus Device software from Synopsys \cite{TCAD}. In the following, we describe the implementation of the sensor geometry in TCAD and the tuning of geometrical parameters by employing C-V measurements and other characterization techniques on the used devices (see section \ref{SS:TCAD_model}).
Once the sensor geometry was optimized, the electric field strength maps were calculated as function of voltage and, the gain of the LGADs was simulated with Sentaurus TCAD using the various impact ionization models available within the tool, as well as the model described by Massey et al.\cite{Massey} (see section \ref{SS:TCAD_simulation}). This procedure is very similar to the works presented in \cite{Mandurrino, Croci_2022, Yang_2021} and  as it was observed in these previous works, we could not find a model that gives a fully satisfying match to our experimental data.
In consequence, we started a model parameter optimization procedure for the three most commonly used models which we label in the following as Massey \cite{Massey}, Overstraeten \cite{VANOVERSTRAETEN1970583}, and Okuto-Crowell \cite{OKUTO1975161}. For the optimization process, the electric field strength maps as function of voltage were extracted from TCAD and used as input in a custom written C++ code for parameter optimization. The TCAD simulations and the C++ code delivered the same LGAD gain values for identical input parameters, with however the C++ code being much faster and therefore convenient for parameter fitting. The so-called gain reduction mechanism \cite{CURRAS2022166530} was carefully avoided in both, the experimental work and the TCAD simulations.  
The overall procedure is described in the following in more detail.

\subsection{Implementation of LGAD device models in TCAD}
\label{SS:TCAD_model}

The LGAD and PIN geometries were implemented in Synopsys TCAD including the JTE and guard ring structures. For the CNM samples the producer provided a complete mask set for all process steps, and information on the process parameters used for the various implantation  and high temperature steps. For the HPK samples only the metal mask layer was available. As for all used device types no sensors with different size, i.e. a different ratio of central pad size to periphery length, were available, it was not possible to correct experimentally for periphery effects by subtracting the C-V curves obtained from different size samples. Therefore, the sensor periphery was implemented in the TCAD simulations to obtain the simulated C-V of the full devices. For all sensor types high resolution optical images were produced at the CERN EP-DT QART lab (using a HIROX RX-2000) to verify metal dimensions and provide input on other structures that were optically visible though their requirement of being based on an etching process. For the HPK samples additional Scanning Electron Microscope – Focused Ion Beam (SEM-FIB) was conducted at the CERN EN-MME lab (using a ZEISS XB540) to reveal the exact position of etches into the oxide layers as well as oxide thickness. The active sensor thickness as well as the bulk doping concentration was extracted from C-V measurements of the PIN and LGAD sensors. Finally, the doping profiles of the front phosphorus implant and the GL boron profile for the CNM LGAD and the HPK2 S1 LGAD were extracted from the SIMS measurements presented in section \ref{SS:SIMS} and fed into the TCAD simulations. The measured depletion voltage of the gain layer ($V_{gl}$) was then used to fine tune the doping concentration of the gain layer in the TCAD simulations. A reduction of the GL boron profile as measured by SIMS by $8\%$ for the HPK2 S1 LGADs and by $5\%$ for the CNM LGADs resulted in a perfect agreement of the $V_{gl}$ in simulation and measurements. The difference between the GL boron concentration as measured by SIMS against the active boron concentration used in the TCAD simulation of a few percent could be attributed to the fact that SIMS measures the total atomic boron concentration while for the electrical device properties only the electrically active boron is relevant, and only part of the implanted boron is activated. On the other hand, the difference might also only reflect the measurement error of the SIMS method itself.
\\
For the HPK2 LGADs of splits 2,3 and 4 no SIMS measurements were available and the GL doping profile was assumed to have the same shape as the one of split 1. This is a very reasonable assumption taking the doping profiles extracted from C-V measurements shown in figure \ref{Figure:Profile_CV}. The peak of the doping profile was found to be at the same position for split 1, 2 and 4, while it was slightly shifted to the surface for split 3. The concentration for all splits was adapted to match the measured $V_{gl}$. The good agreement between the measured C-V curves and the C-V curves simulated in TCAD, can be seen in figure \ref{Figure:CV} where both, measured and simulated data, are displayed.
\subsection{TCAD simulations of electric field profiles and gain curves}
\label{SS:TCAD_simulation}
\begin{figure}[thb]
\centering
\hfil
\includegraphics[width=0.8\columnwidth]{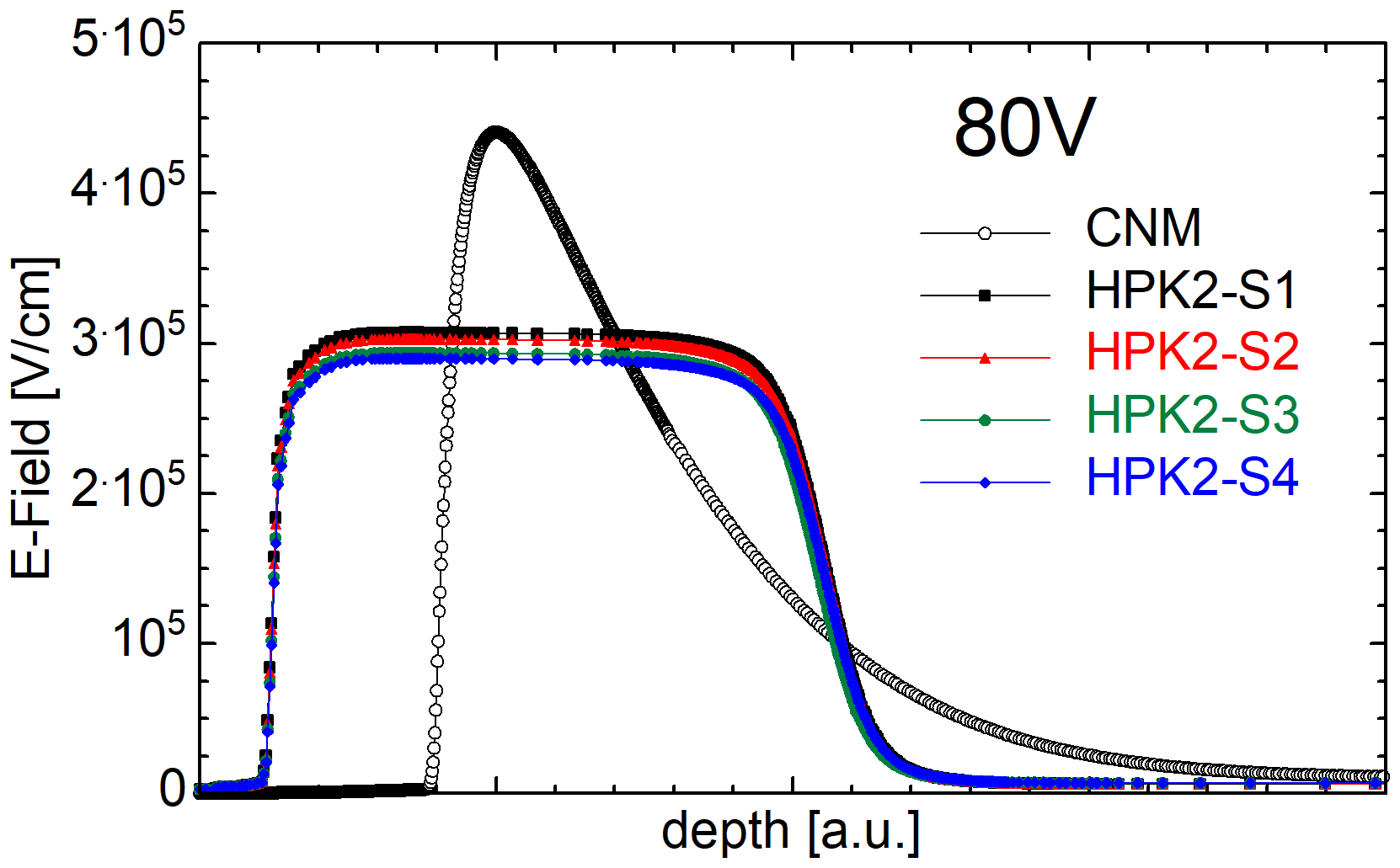}
\hfil
\caption{Simulated electric field strength profiles $E(x)$ at the same bias voltage ($80\,V$) for the five studied LGAD types.} 
\label{Figure:EField_LGADs}
\centering
\end{figure}

The TCAD model described in the previous section resulted in detailed electric field profiles. The experimental gain measurements were obtained with the TCT technique by shining the laser onto a central part of the sensor. Therefore, at the measurement position the electric field (and the gain) is not influenced by the LGAD periphery and can be simulated with a much simpler 1D geometry. 3D and 1D simulations of the electric field and the gain in the measurement position showed no difference. As an example, depth profiles of the electric field strength for all five LGAD types at a voltage of $80\,V$ are shown in figure \ref{Figure:EField_LGADs}.\\ 

The TCAD simulation of gain curves as function of voltage and temperature, followed the same procedure as the experimental work. The gain for a given voltage was obtained by dividing the simulation result for the LGAD by the result for the corresponding PIN sensor extracted above full depletion. The gain was determined by either simulating time resolved TCT measurements or by simulating steady state leakage current with a homogeneous SRH generation across the bulk of the device. Both methods yield the same gain values for voltages above device depletion and as long as the regime for gain reduction effects is avoided \cite{CURRAS2022166530}.\\ 

\begin{figure}[thb]
\centering
\includegraphics[width=0.75\columnwidth]{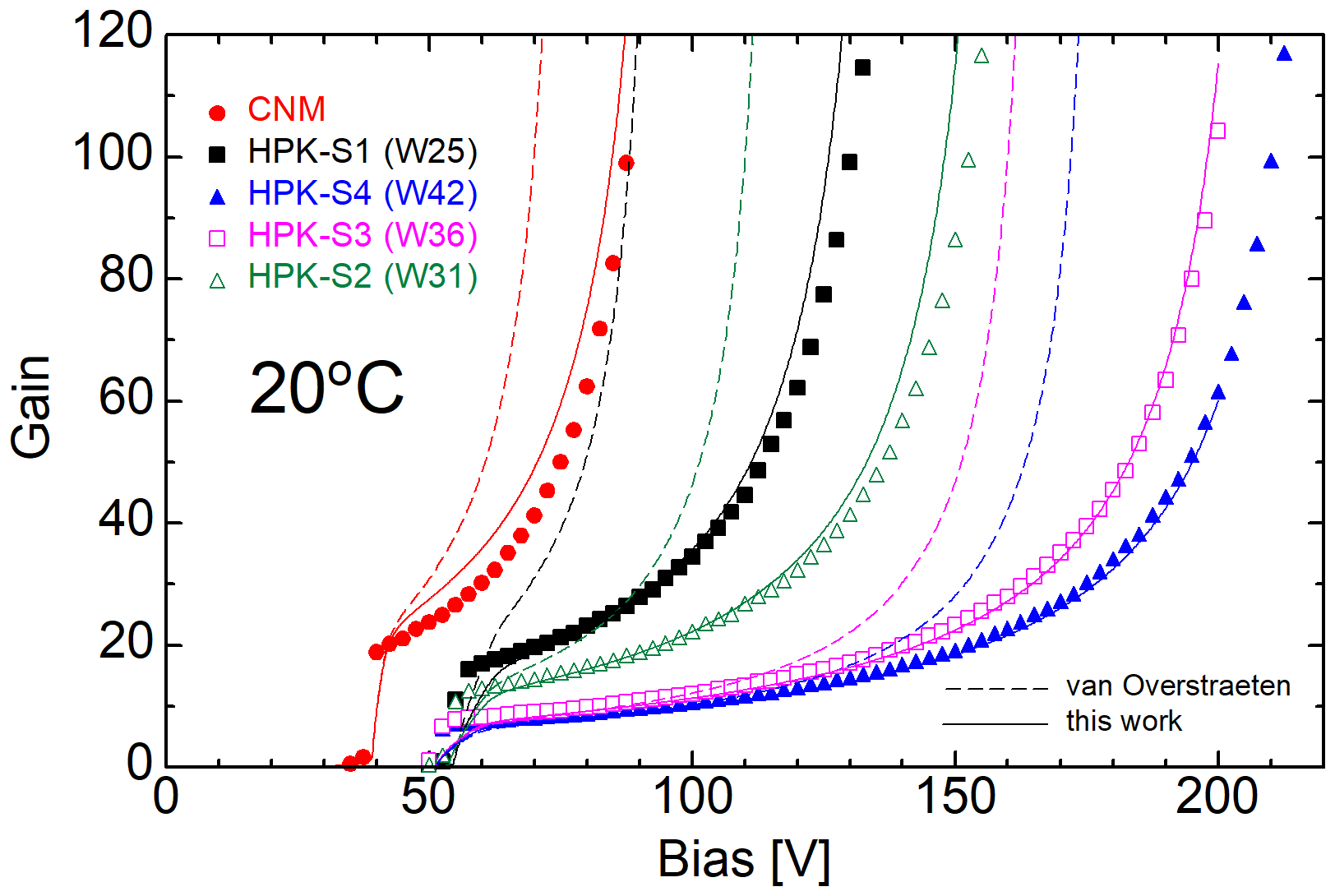}\hfil
\caption{Disagreement between measured and simulated gain using the Overstraeten model with default parameters for all LGAD devices studied in this work (data taken at $20\,^oC$). Measured data are given by the symbols. The TCAD simulations are shown as lines: dashed lines for the model with default parameters and solid lines for the model with optimized input parameters.} 
\label{Figure:gain-20C-all}
\centering
\end{figure}

\begin{figure}[thb]
\centering
\includegraphics[width=0.5\columnwidth]{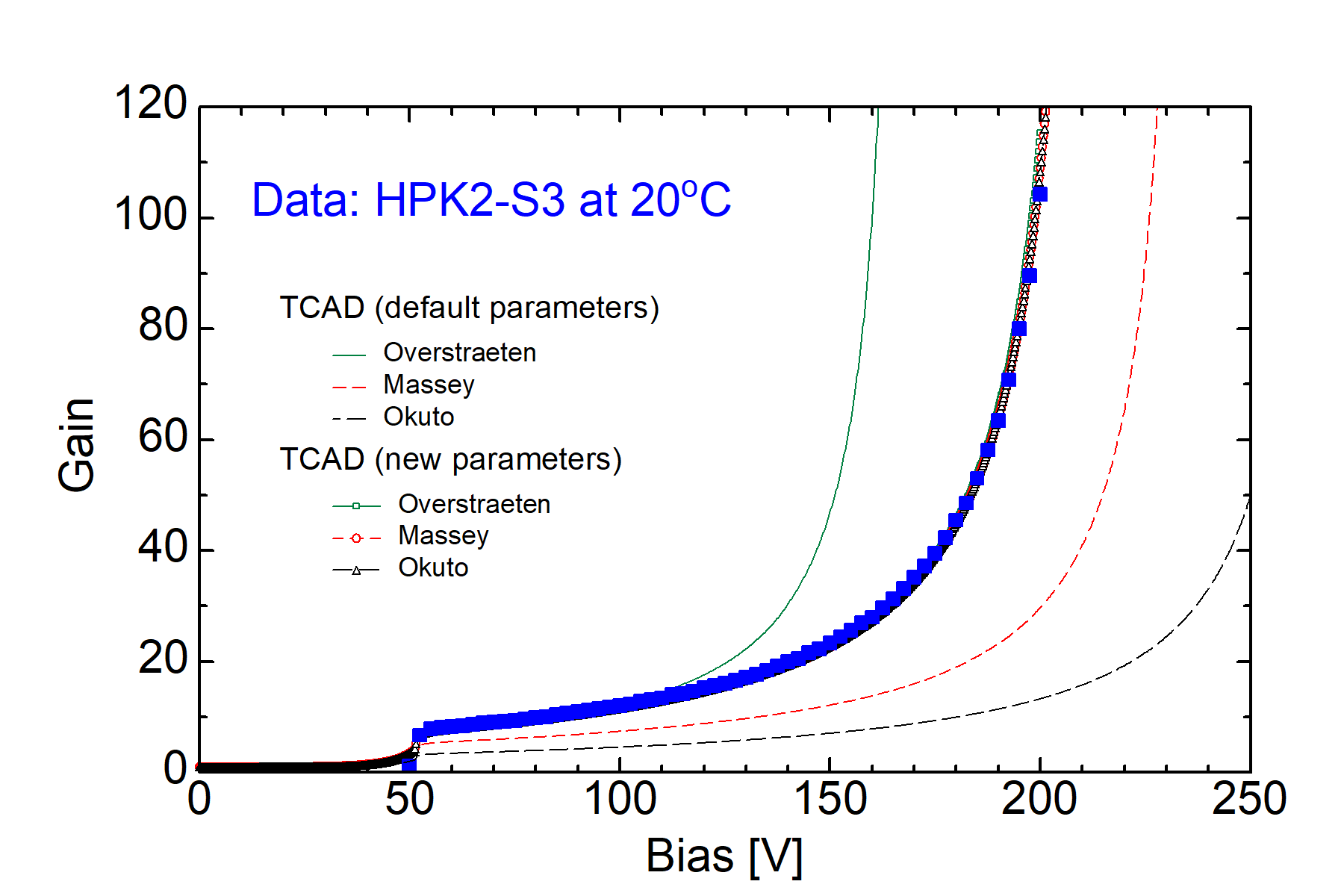}\hfil
\includegraphics[width=0.5\columnwidth]{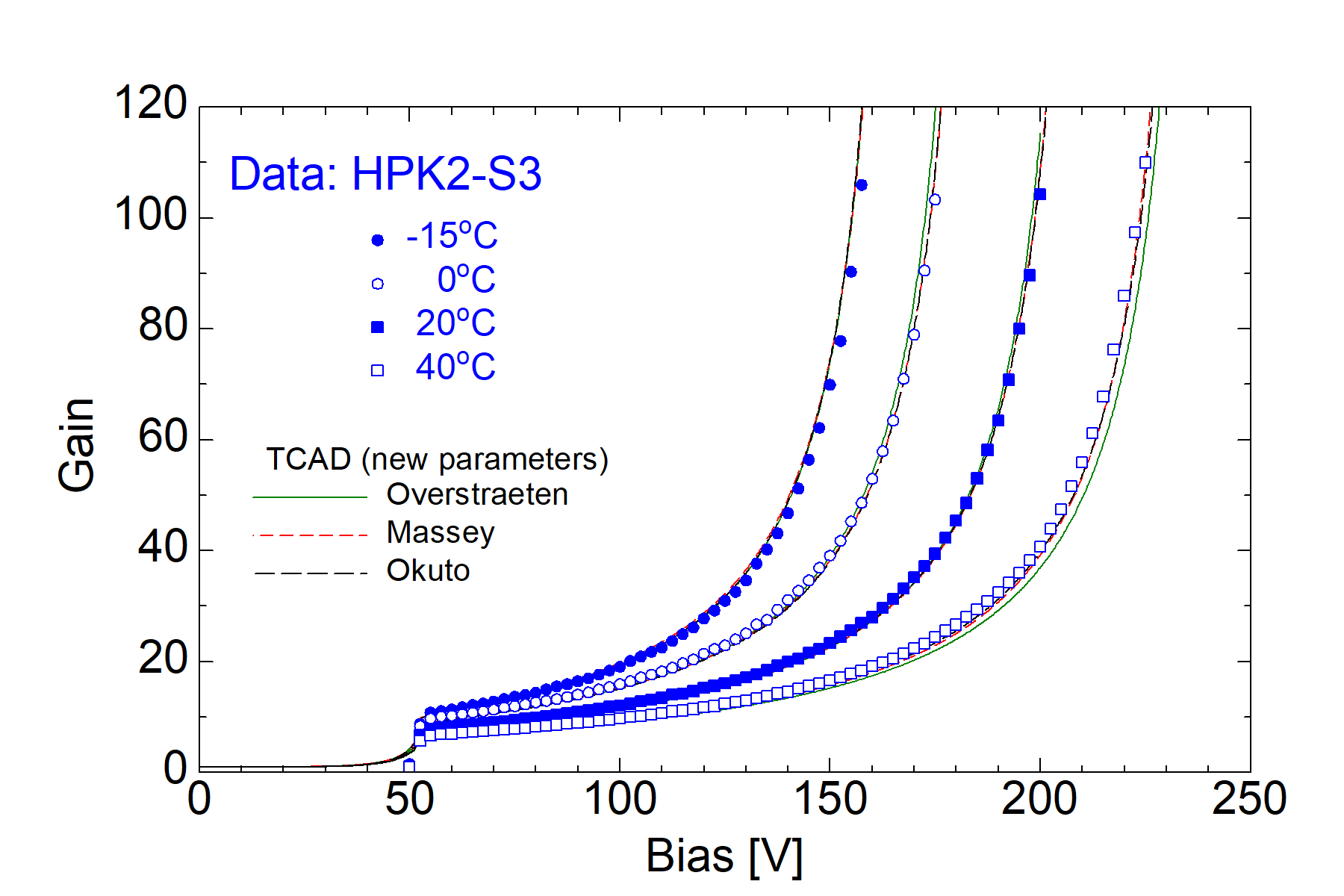}\hfil
\caption{(left) Measured gain values for the HPK2-S3 device type at $20^oC$ (symbols), and TCAD simulations for three different impact ionization models with default parameters (lines) and optimized parameters (lines with symbols). The optimized parameters give a good agreement, while the default parameters give deviations. (right) Simulations with optimized parameters for the three indicated models at various temperatures.} 
\label{Figure:gain-W36-T}
\centering
\end{figure}
All impact ionization models available in the Sentaurus software and the Massey model were tested using the default parameter sets, i.e. the published parameter sets. Figure \ref{Figure:gain-20C-all} gives a comparison of the experimental and simulated gain curves for the Overstraeten model. For better visibility only data at $20^\circ\,C$ are presented. It is clearly visible that the Overstraeten model with default parameters gives higher gain values than measured. In Figure \ref{Figure:gain-W36-T}, data for the LGAD type HPK2-S3 are shown. For the data at $20^oC$ (Figure \ref{Figure:gain-W36-T}, left) the Massey and Okuto-Crowell models underestimate the gain while the Overstraeten model overestimates the gain.  
This disagreement motivated us to optimize the parameters of these models to get a better agreement between measured and simulated data, as demonstrated in Figure \ref{Figure:gain-W36-T} (right) and described in more detail below.
For a faster optimization of the parameters, the impact ionization equation was implemented in C++ and by the end of the process, the models with the optimized parameters were tested in TCAD (see e.g. figures \ref{Figure:gain-20C-all} and \ref{Figure:gain-W36-T}). Both simulation methods: TCAD and C++, gave the same gain.

\subsection{Optimization procedure of impact ionization parameters}
The impact ionization model parameters were adjusted by a custom written C++ program to match our experimental data. The program allowed to calculate the LGAD gain as function of voltage and temperature based on the input of the electric field strength profile from the TCAD simulations. The calculations are based on the solution of the differential equation for electron ($J_n$) and hole ($J_p$) current densities under impact ionization \cite{McIntyre1999} 
\begin{equation}
\label{diff_eq}
\frac{dJ_p}{dx} = \frac{-dJ_n}{dx} = \alpha_nJ_n + \alpha_pJ_p + g(x)
\end{equation}
with $\alpha_n$ and $\alpha_p$ being the impact ionization coefficients (in units of $cm^{-1}$) for electrons and holes respectively, and $g(x)$ describing the generation of excess charge carriers. The solution is given by 
\begin{equation}
\label{sol_eq}
J = J_n(x) + J_p(x) = M_nJ_n(d) + M_pJ_p(0) + \int_{0}^{d} g(x)M(x)dx,
\end{equation}
with
\begin{equation}
\label{gain_eq}
M(x) = \frac{exp\left (-\int_{x}^{d}(\alpha_n-\alpha_p)d\eta\right )}{1-\int_{x}^{d}\alpha_n\:exp\left (-\int_{\xi}^{d}(\alpha_n-\alpha_p)d\eta\right )d\xi},
\end{equation}
and $M_n = M(x=d)$ being the multiplication of electrons injected from the back-electrode at $x=d$ and $M_p = M(x=0)$ the multiplication of holes injected from the front-electrode at $x=0$. As the charge in our case is generated by the laser illumination throughout the bulk, only the last term in equation \ref{sol_eq} is relevant and was taken into account in the numerical calculations.\\ Three different models for the impact ionization coefficients $\alpha_{n,p}$ were implemented and optimized for their parameters. All three models are following the Chynoweth law \cite{Chynoweth}, but each  uses a different formalism for the parameterization: 
\begin{itemize}
\item
In the {\bf Massey model} \cite{Massey} the $\alpha_{n,p}$ coefficients follow the equation
\begin{equation}
\label{Massey_eq}
\alpha_{n,p}(E,T) = A_{n,p} \: exp\left (-\frac{C_n+D_{n,p}\cdot T}{E}\right ),
\end{equation}
with $E$ being the electric field strength, $T$ the absolute temperature and $A_{n,p}$, $C_{n,p}$ and $D_{n,p}$ the six model parameters for the two impact ionization coefficients.
\item
In the {\bf Van Overstraeten - de Man model} \cite{VANOVERSTRAETEN1970583}, the $\alpha_{n,p}$ are described by
\begin{equation}
\label{Overst1_eq}
\alpha_{n,p}(E,T) = \gamma\: A_{n,p} \: exp\left (-\gamma \: \frac{B_{n,p}}{E}\right ),
\end{equation}
\begin{equation}
\label{Overst2_eq}
\text{with}\quad \gamma = \frac{tanh \left (  \frac{\hbar\omega_{op}}{2kT_0}  \right )}{tanh \left (  \frac{\hbar\omega_{op}}{2kT}  \right )},
\end{equation}
where $k$ is the Boltzmann constant, $T_0=300\,K$ is the reference temperature, the parameter $\hbar\omega_{op}$ stands for the optical phonon energy and $A_{n,p}$ and $B_{n,p}$ are further parameters.
\item In the {\bf Okuto-Crowell model} \cite{OKUTO1975161}, the $\alpha_{n,p}$ coefficients follow the equation:
\begin{equation}
\label{Okuto_eq}
\alpha_{n,p}(E,T) = A_{n,p}(1+C_{n,p}\cdot \tilde{T})E \: exp \left (- \left (\frac{B_{n,p}(1+D_{n,p}\cdot \tilde{T})}{E}\right )^2 \right ),
\end{equation}
with $\tilde{T} = T-300\,K$ and eight parameters $A_{n,p}$,$B_{n,p}$, $C_{n,p}$ and $D_{n,p}$.
\end{itemize}
The original values (i.e. the published values and default values in Synopsys TCAD) for the different parameters in the equations above are listed  in the appendix (section \ref{S:9}) in table\,\ref{table_2} for Massey, table \ref{table_3} for Overstraeten - de Man and \ref{table_4} for Okuto-Crowell. For the Overstraeten model, only the values for the low electric field region ($E\,<\,\num{4.0e5}\,V cm^{-1}$) are given, as in our parameterization we used only one set of parameters and not two as in the original publication.\\
The optimization of the parameters was done using the least squares method and therefore the sum of the squares of the normalized residuals was minimized for each model. The residuals were defined as the difference between the measured gain and the value provided by the model, divided by the measured gain.

\section{Optimized parameters for impact ionisation models}
\label{S:Results}
The full experimental data set of gain measurements for the 5 LGAD types up to a gain of 100 and in the temperature range from $-15\,^oC$ to $40\,^oC$, was subjected to the optimization procedure described in the section \ref{S:Method} in a single fit for each model.
The optimal values for the different parameters in the equations can be found in the appendix section (section \ref{S:9}) in tables\,\ref{table_2},\,\ref{table_3} and\,\ref{table_4}. With the default parameterization, the result of the sum of the residuals was 30.4 for Massey (underestimates the gain), \num{5.39e4} for Overstraeten (overestimates the gain) and 38.3 for Okuto-Crowell (underestimates the gain). After the optimization of the parameters, the sum of the residuals was reduced to: 0.515 for Massey, 0.681 for Overstraeten and 0.556 for Okuto-Crowell. With all three models the simulated gain was in good agreement with the measured gain in all the LGADs across the full experimental data range. The maximum simulated electric field was \num{4.5e5}\,V$cm^{-1}$ and the lowest electric field, for which we had sensitivity, was \num{2.8e5}\,V$cm^{-1}$. A detailed comparison between experimental data and the gain simulations using the models with the optimized parameters is given in the following.
\subsection{Electric field dependence of impact ionization}
\label{SS:4_1}
\begin{figure}[t]
\centering
\includegraphics[width=0.33\columnwidth]{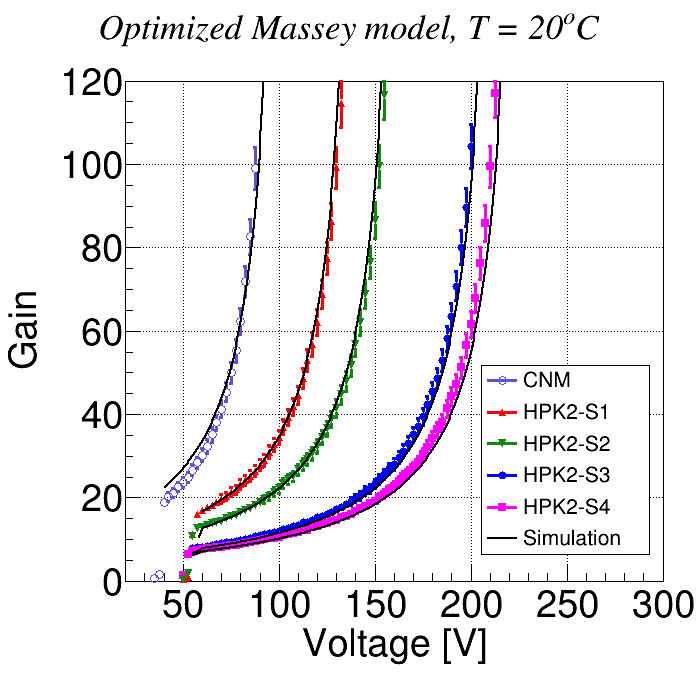}\hfil
\includegraphics[width=0.33\columnwidth]{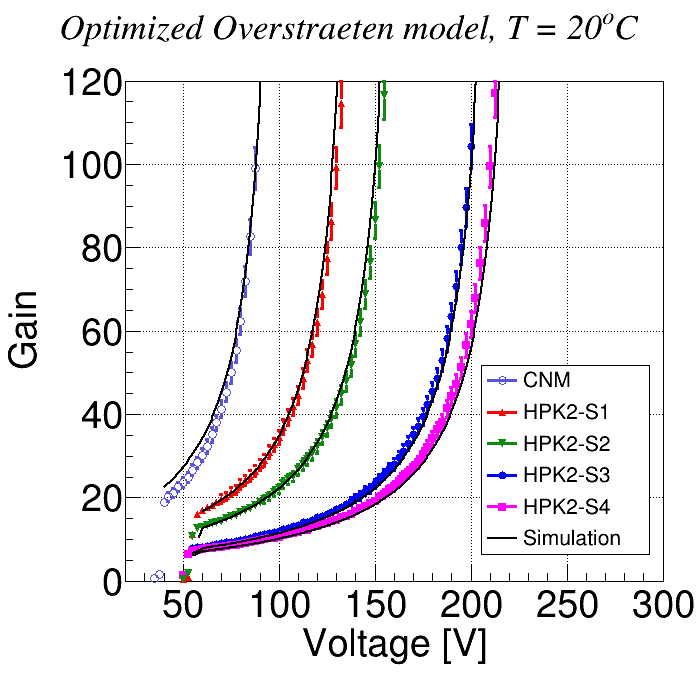}\hfil
\includegraphics[width=0.33\columnwidth]{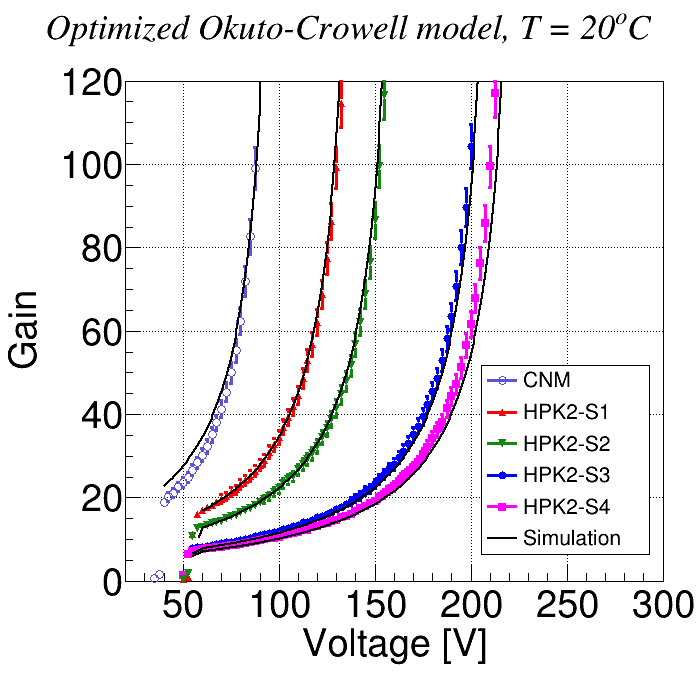}\hfil
\caption{Agreement between the measured and simulated gain, at $20\,^oC$, after the optimization of the parameters for the three models indicated in the figure titles.} 
\label{Figure_Gain_20C}
\centering
\end{figure}

\begin{figure}[t]
\centering
\includegraphics[width=0.48\columnwidth]{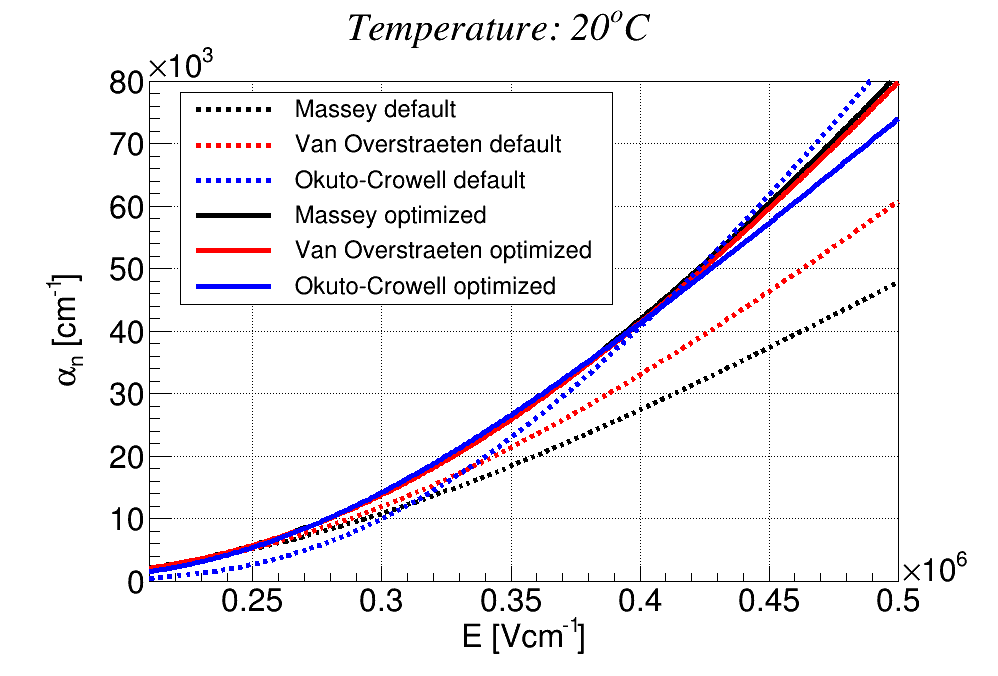}\hfil
\includegraphics[width=0.48\columnwidth]{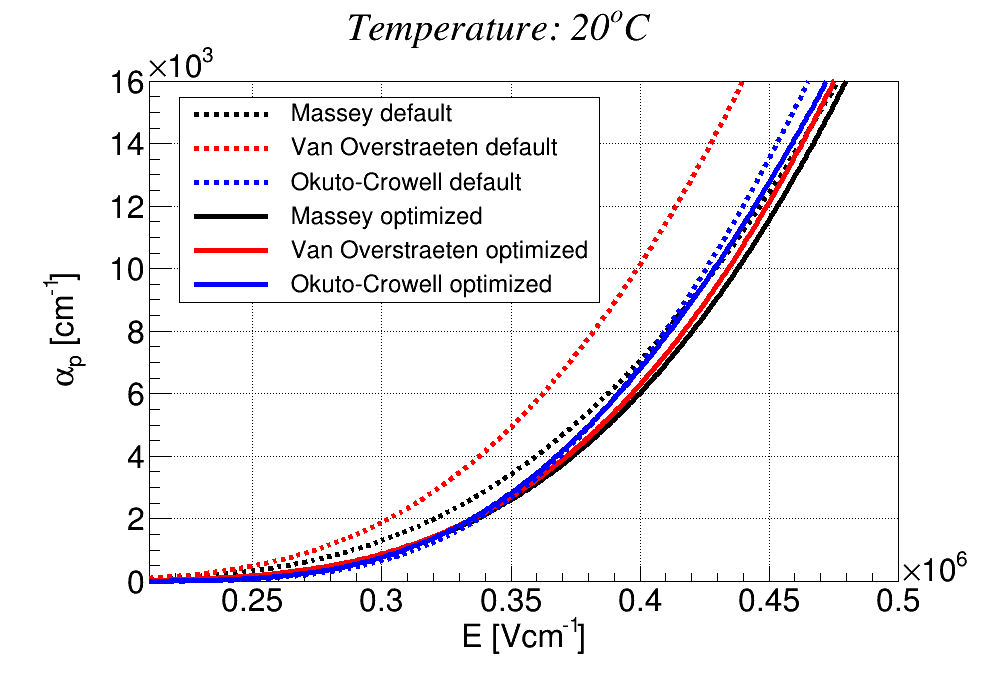}\hfil
\caption{Impact ionization coefficients as a function of the electric field at $20\,^oC$. Coefficients based on the original model values (dashed lines) and the optimized parameters are shown.} 
\label{Figure_alpha_vs_E}
\centering
\end{figure}

The agreement between the model with the new parameterization, as obtained with the C++ code, and the measured gain at $20\,^oC$ is shown in figure \ref{Figure_Gain_20C}. After the optimization of the parameters, all three models converge to almost identical solutions. The new $\alpha_{n}(E)$ and $\alpha_{p}(E)$ coefficients as a function of the electric field at $20\,^oC$ are shown in figure \ref{Figure_alpha_vs_E}. All models give very similar impact ionization coefficients as a function of the electric field and therefore, the simulated gain is almost the same. This was not the case when using the default parameterization (see figure \ref{Figure:gain-W36-T}). The Overstraeten model with the default parameterization is overestimating the gain because $\alpha_{p}(E)$ is too high, while the Massey and Okuto-Crowell models are underestimating the gain because $\alpha_{n}(E)$ is too low.

\subsection{Temperature dependence of the impact ionization}
\label{SS:4_2}
\begin{figure}[t]
\centering
\includegraphics[width=0.33\columnwidth]{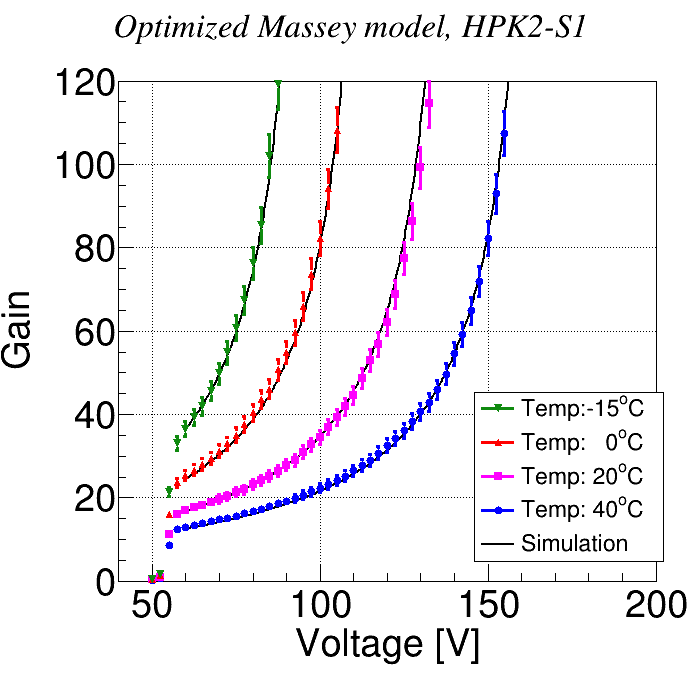}\hfil
\includegraphics[width=0.33\columnwidth]{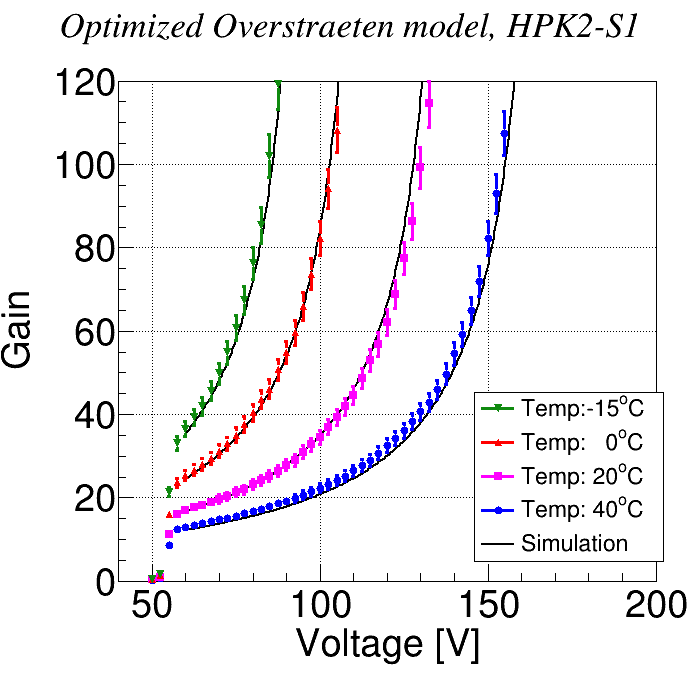}\hfil
\includegraphics[width=0.33\columnwidth]{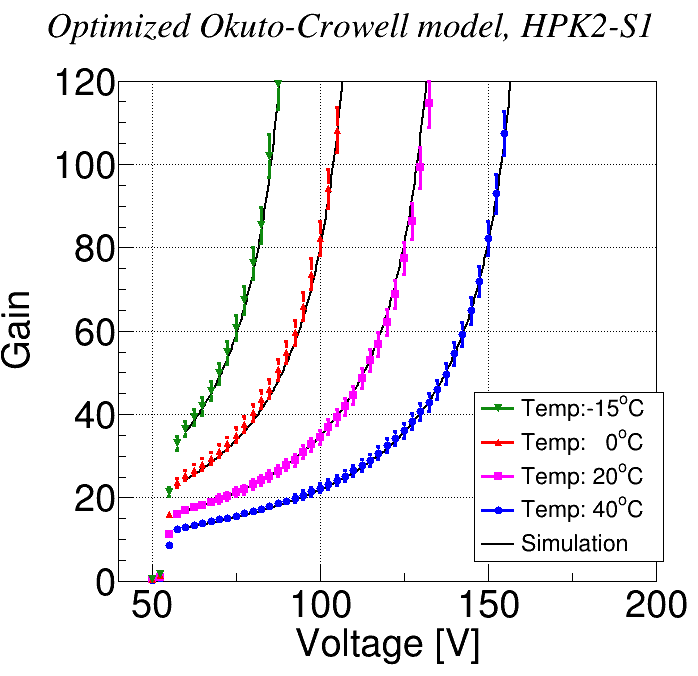}\hfil
\caption{Measured and simulated gain of the HPK2-S1 LGAD at different temperatures, after the optimization of the parameters for the three models indicated in the titles of the three plots.} 
\label{Figure_10}
\centering
\end{figure}

\begin{figure}[t]
\centering
\includegraphics[width=0.48\columnwidth]{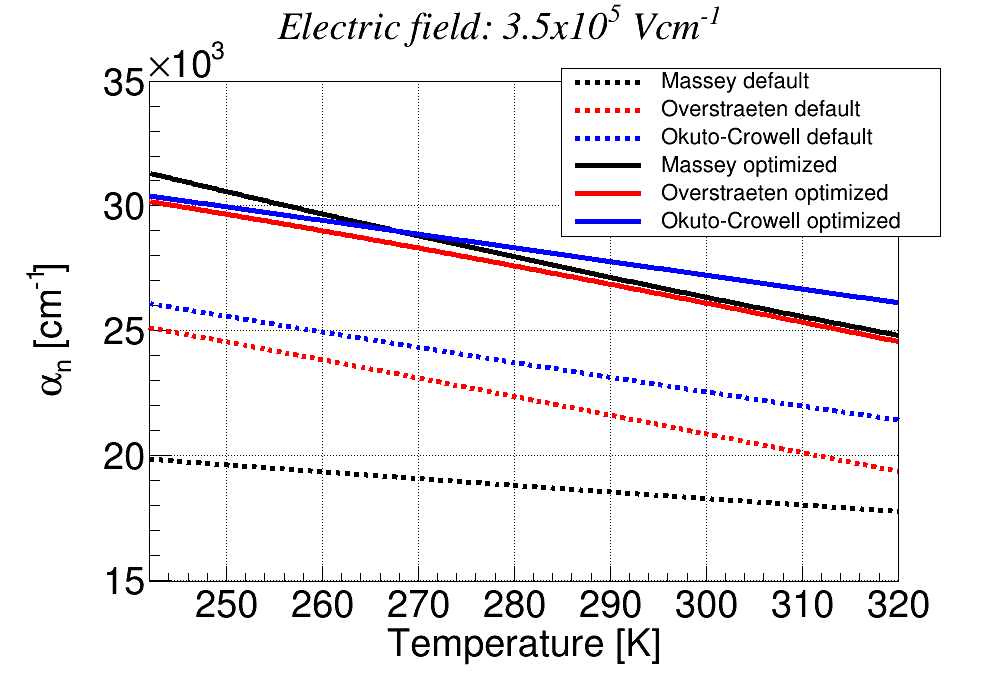}\hfil
\includegraphics[width=0.48\columnwidth]{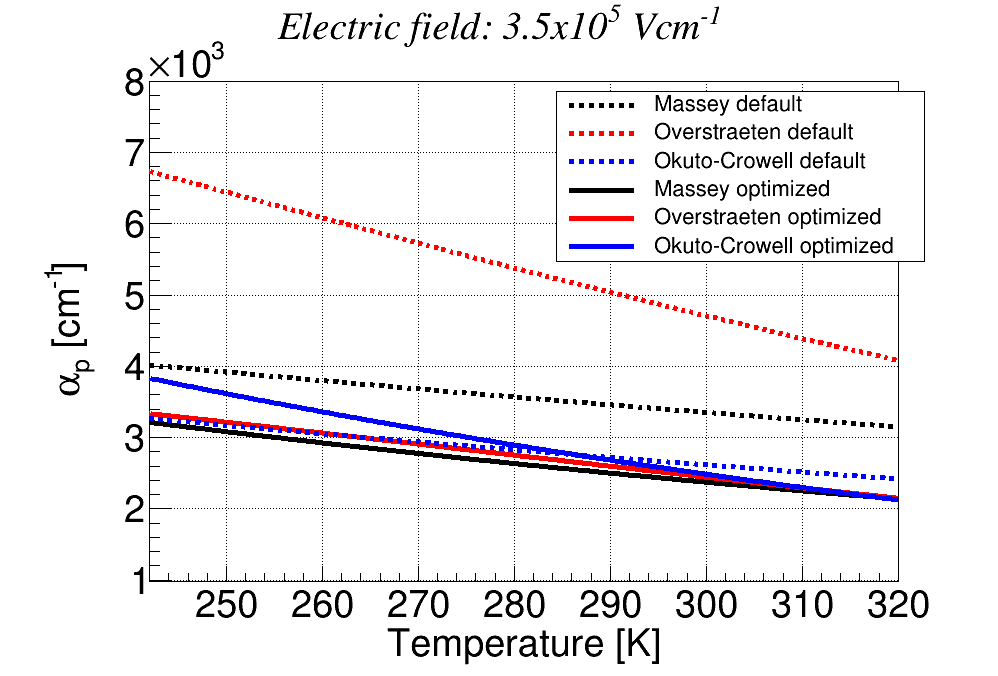}\hfil
\caption{New impact ionization coefficients as a function of the temperature for $E = \num{3.5e5}\,V cm^{-1}$. In the plot are included the coefficients by default for the three methods too.} 
\label{Figure_11}
\centering
\end{figure}

When the gain is simulated at a temperature other than $20\,^oC$, the disagreement between the measured gain and the one simulated with the original parameters is even more significant than the one shown in figure\,\ref{Figure:gain-20C-all} for $20\,^oC$. With the new parameters set, optimized against the full set of experimental data taken for the five LGAD types between $-15\,^oC$ and $+40\,^oC$, a good agreement between simulated and measured data is found for all temperatures. As example, the results for the HPK2-S1 LGAD type are shown in figure\,\ref{Figure_10}. As shown for the electric field strength dependence in the previous section, there is not a significant difference between the three impact ionization models with optimized parameter sets in the investigated temperature range. The new $\alpha_{n}(T)$ and $\alpha_{p}(T)$ coefficients as a function of the temperature for an electric field of  $\num{3.5e5}\,V cm^{-1}$ are shown in figure \ref{Figure_11}.
The three original models were underestimating $\alpha_{n}(T)$, while $\alpha_{p}(T)$ was overestimated by Massey and Overstraeten, but not by Okuto-Crowell with respect to our data. Also, despite these three models using very different formulation for the temperature dependence of the impact ionization, the $\alpha_{n}(T)$ and $\alpha_{p}(T)$ coefficients obtained after the optimization of the parameters are in good agreement. From the rising deviation of the three models towards the lowest and highest temperature investigated, differences are expected outside the temperature window studied here, indicating that any projection towards lower or higher temperature should be treated with great care.

\subsection{Error discussion}
\label{SS:4_2}

To understand the validity of the new parameterization, it is important to understand the sources and influence of the errors in the method. Taking as an example the HPK2-S1 LGAD, we show how the errors in the input data can strongly modify the simulated gain and therefore give different values for the impact ionization coefficients.

\begin{itemize}

\item A potential source of error lies in the experimental measurement of the gain. Assuming that there is no uncertainty in the reverse bias voltage and temperature, the main uncertainty arises from the already mentioned gain reduction mechanism \cite{CURRAS2022166530}. For the HPK2-S1 LGAD we obtain a gain of 25 at $80\,V$ and $20\,^oC$ when the gain reduction is negligible, but if the laser intensity is increased to 10 MIPs (x10 charge density) the measured gain will drop to a value of 17. This effect will be even stronger at higher gain values. Therefore, such an error will lead to a model that will underestimate the gain.

\item The doping profile of the GL that is needed to generate the electric field map in the TCAD device model is also a critical input. In this regard, an error in the active boron concentration or in the position of the GL doping profile, will lead to a modification in the electric field profile and therefore, will affect the parametrization of the models. For example, a decrease of $2\%$ in the concentration of the GL or a shift of the GL position by $200\,nm$ closer to the front electrode, will drop the simulated gain from a value of 25 (at 80 V and $20\,^oC$) to a value of 17. An uncertainty in any of these two parameters will lead to a model that, depending on the direction of the error, underestimates or overestimates the gain.

\item The final boron concentration profile and position of the GL used to simulate the electric field profile, should ideally be adjusted using the measured capacitance curve of the LGADs, which can be reliably measured and simulated and thus, used to minimize the error introduced by these two parameters. However, this requires an exact knowledge on the LGAD geometrical structure, including the device periphery, e.g. JTE and guard rings. A doping profiling extracted from C-V curves assuming a parallel plate sensor geometry with fixed area $A$ (see section \ref{SS:Electrical_characterization}) is not sufficient in this sense. 

\item Discretization errors in the simulations can also be a source of errors, in particular for the calculation of the gain, and should be investigated by variation of the number of nodal points. In our case, custom C++ and TCAD device simulations yielded the same gain curves giving confidence in not being impacted by this error.

\end{itemize}

If we take into account all these possible uncertainties, the error estimation in the modelling of the electric field and therefore, in the simulated gain is complicated. Thus, an approximation on the uncertainty in the new parameterization can be extracted from the dispersion in the impact ionization parameters obtained by the three different models studied. In this way, we can conclude that the error in the impact ionization parameters should be around $10\%$ in the range of electric fields and temperatures studied. For the experimental measurement of the gain, the measured error was $\sim5\%$.
\section{Summary}
\label{S:Summary}
We present a new parameterization of impact ionization coefficients in silicon devices. The parameterization is derived from a large set of pulsed IR-laser based gain measurements on 5 different types of LGADs including devices with deep and shallow gain layers. New parameters are provided for the three most commonly used models in the LGAD community: Massey, Overstraeten and Okuto-Crowell.\\
The original parameterization of these models did not provide a satisfying agreement with our experimental data motivating the presented work. TCAD device models were conceived for all LGADs based on the available knowledge about the used devices including mask files, processing information, doping profiles measured with SIMS, as well as Transient Current Technique (TCT) and capacitance voltage (CV) measurements. The TCAD simulations yielded the electric field strength maps that were then used in a custom written software to optimize the parameters in the three different impact ionization coefficients models.
After the parameter optimization, all three models give very similar solutions for the gain in very good agreement with the measured data. We estimate an error of $\sim10\%$ in the parametrization obtained and its validity covers the range in temperatures from $-15\,^oC$ to $40\,^oC$, and a range in electric fields from $\num{2.8e5}\,V cm^{-1}$ to $\num{4.5e5}\,V cm^{-1}$, and a maximum gain of 100 for the used LGAD devices.

\section{Acknowledgements}
This work has been performed in the framework of the RD50 collaboration and the CERN EP R\&D programme on technologies for future experiments. The authors like to thank Salvador Hidalgo from CNM, Barcelona, Spain for the detailed information provided on the LGADs produced at CNM.

\newpage

\section{Appendix}
\label{S:9}

\begin{table}[H]
\begin{center}
\begin{tabular}{c|cr|cr}
\cline{2-5}
\multicolumn{1}{l|}{} & \multicolumn{2}{c|}{Massey default} & \multicolumn{2}{c|}{Massey optimized}  \\ \hline
\multicolumn{1}{|c|}{Parameter}     & \multicolumn{1}{c|}{electrons}                & \multicolumn{1}{c|}{holes}    & \multicolumn{1}{c|}{electrons}  & \multicolumn{1}{c|}{holes}\\ \hline
\multicolumn{1}{|l|}{$A\,(cm^{-1}$) } & \multicolumn{1}{r|}{\num{4.43e5}}  & \multicolumn{1}{r|}{\num{1.13e6}}  & \multicolumn{1}{r|}{\num{1.186e+06}} & \multicolumn{1}{r|}{\num{2.250e+06}} \\ \hline
\multicolumn{1}{|l|}{$C\,(V cm^{-1}$)} & \multicolumn{1}{r|}{\num{9.66e5}}  & \multicolumn{1}{r|}{\num{1.71e6}}  & \multicolumn{1}{c|}{\num{1.020e+06}}      & \multicolumn{1}{r|}{\num{1.851e+06}}   \\ \hline
\multicolumn{1}{|l|}{$D\,(V cm^{-1}K^{-1}$)} & \multicolumn{1}{r|}{\num{4.99e2}}  & \multicolumn{1}{r|}{\num{ 1.09e3}} & \multicolumn{1}{c|}{\num{1.043e+03}}      & \multicolumn{1}{r|}{\num{1.828e+03}} \\ \hline
\end{tabular}
\caption{Ionization parameters for the Massey model. The default values are on the left side of the table and the optimized values from this paper on the right.}
\label{table_2}
\end{center}
\end{table}

\begin{table}[H]
\begin{center}
\begin{tabular}{c|cr|cr}
\cline{2-5}
\multicolumn{1}{l|}{} & \multicolumn{2}{c|}{ Overstraeten default} & \multicolumn{2}{c|}{ Overstraeten optimized}  \\ \hline
\multicolumn{1}{|c|}{Parameter}     & \multicolumn{1}{c|}{electrons}                & \multicolumn{1}{c|}{holes}    & \multicolumn{1}{c|}{electrons}  & \multicolumn{1}{c|}{holes}\\ \hline
\multicolumn{1}{|l|}{$A\,(cm^{-1}$) } & \multicolumn{1}{r|}{\num{7.03e5}}  & \multicolumn{1}{r|}{\num{1.582e6}}  & \multicolumn{1}{r|}{\num{1.149e+06}} & \multicolumn{1}{r|}{\num{2.519e+06}} \\ \hline
\multicolumn{1}{|l|}{$B\,(V cm^{-1}$)} & \multicolumn{1}{r|}{\num{1.231e6}}  & \multicolumn{1}{r|}{\num{2.036e6}}  & \multicolumn{1}{c|}{\num{1.325e+06}}      & \multicolumn{1}{r|}{\num{2.428e+06}}   \\ \hline
\multicolumn{1}{|l|}{$\hbar\omega_{op}\,(eV)$} & \multicolumn{1}{r|}{\num{0.063}}  & \multicolumn{1}{r|}{\num{0.063}} & \multicolumn{1}{r|}{\num{0.0758}}      & \multicolumn{1}{r|}{\num{0.0758}} \\ \hline
\end{tabular}
\caption{Ionization parameters for the Overstraeten model. The default values are on the left side of the table and the optimized values from this paper on the right. The parameters shown in the table correspond to the low electric field ones: $E < \num{4.0e5}\,V cm^{-1}$.}
\label{table_3}
\end{center}
\end{table}

\begin{table}[H]
\begin{center}
\begin{tabular}{c|cr|cr}
\cline{2-5}
\multicolumn{1}{l|}{} & \multicolumn{2}{c|}{Okuto-Crowell default} & \multicolumn{2}{c|}{Okuto-Crowell optimized}  \\ \hline
\multicolumn{1}{|c|}{Parameter}     & \multicolumn{1}{c|}{electrons}                & \multicolumn{1}{c|}{holes}    & \multicolumn{1}{c|}{electrons}  & \multicolumn{1}{c|}{holes}\\ \hline
\multicolumn{1}{|l|}{$A\,(V^{-1}$) } & \multicolumn{1}{r|}{\num{0.426}}  & \multicolumn{1}{r|}{\num{0.243}}  & \multicolumn{1}{r|}{\num{0.289}} & \multicolumn{1}{r|}{\num{0.202}} \\ \hline
\multicolumn{1}{|l|}{$B\,(Vcm^{-1}$)} & \multicolumn{1}{r|}{\num{4.81e5}}  & \multicolumn{1}{r|}{\num{6.53e5}}  & \multicolumn{1}{c|}{\num{4.01e5}}      & \multicolumn{1}{r|}{\num{6.40e5}}   \\ \hline
\multicolumn{1}{|l|}{$C\,(K^{-1}$)} & \multicolumn{1}{r|}{\num{3.05e-4}}  & \multicolumn{1}{r|}{\num{5.35e-4}} & \multicolumn{1}{c|}{\num{9.03e-4}}      & \multicolumn{1}{r|}{\num{-2.20e-3}} \\ \hline
\multicolumn{1}{|l|}{$D\,(K^{-1}$)} & \multicolumn{1}{r|}{\num{6.86e-4}}  & \multicolumn{1}{r|}{\num{5.67e-4}} & \multicolumn{1}{c|}{\num{1.11e-3}}      & \multicolumn{1}{r|}{\num{8.25e-4}} \\ \hline
\end{tabular}
\caption{Ionization parameters for the Okuto-Crowell model. The default values are on the left side of the table and the optimized values from this paper on the right.}
\label{table_4}
\end{center}
\end{table}

\newpage




\bibliographystyle{model1-num-names}
\bibliography{bibliography.bib}






\end{document}